\documentclass[epj]{bcjour}
\usepackage{graphics}
\usepackage{epsfig}
  \def\dm{\Delta m^2}
\def\dtm{\Delta \tilde{m}} 
\begin{document}

 
%
\title{Prospects of measuring $\sin^2 2\Theta_{13}$ and 
the sign of $\Delta m^2$ with a massive magnetized detector for 
atmospheric neutrinos}
\author{
       T.~Tabarelli de Fatis
       }  

\institute{INFN, Sezione di Milano, Piazza della Scienza 3, 
                 I-20126 Milan, Italy}
%
\date{February 22, 2002}
%
\abstract{The pattern of oscillation parameters emerging from current 
          experimental data can be further elucidated by the
          observation of matter effects. In contrast to planned experiments
          with conventional neutrino beams, atmospheric neutrinos offer
          the possibility to search for Earth-induced matter effects
          with very long baselines. Resonant matter effects are
          asymmetric on neutrinos and anti-neutrinos, depending on the
          sign of $\dm$. In a three-generation oscillation scenario, 
          this gives access to the mass hierarchy of neutrinos, while
          the size of the asymmetry would measure the admixture of
          electron neutrinos to muon/tau neutrino oscillations (the  
          mixing angle $\Theta_{13}$). 
          The sensitivity to these effects is discussed after the
          detailed simulation of a realistic experiment based on a
          massive detector for atmospheric neutrinos with charge
          identification. We show how a detector, which measure and
          distinguish between $\nu_\mu$ and $\overline{\nu}_\mu$
          charged current events, might be sensitive to matter
          effects using atmospheric neutrinos, provided the mixing
          angle $\Theta_{13}$ is large enough. 
%
} 
\titlerunning{Prospects of measuring $\sin^2 2\Theta_{13}$ and 
the sign of $\Delta m^2$ ...}
\maketitle

\section{Introduction}
\label{intro}

The very long baselines available with atmospheric neutrinos offer 
the possibility to search for Earth-induced matter effects. 
In that endeavour, atmospheric neutrino experiments are not 
contested by current and planned accelerator beam experiments -- 
whose baselines are too short for a significant effect -- and 
offer the opportunity to test the neutrino mass hierarchy.
The observation of matter effects might also be possible with upgraded
conventional neutrino beams \cite{Barger} and at neutrino
factories \cite{nufact}. 

Matter effects can play an important role if there are significant
contributions of $\nu_e$ or $\nu_{\rm sterile}$ to atmospheric
neutrino oscillations. The non-observation of large matter effects has
been already exploited by Super-Kamiokande to exclude a large
contribution of $\nu_\mu - \nu_{\rm sterile}$ oscillations to
atmospheric neutrinos \cite{SK,SK-2001}.  For a contribution of
non-maximal $\nu_\mu - \nu_{\rm sterile}$ oscillations, matter effects
would also manifest themselves in differences in the oscillation
patterns for neutrinos and anti-neutrinos. These differences could be
measured with a magnetized detector with the capability to identify
the muon charge \cite{Monolith}.

Matter effects could be detectable even in standard three flavour
oscillation scenarios \cite{Banul}.  A sub-dominant $\nu_e$ mixing
could sizeably modify the $\nu_\mu$ transition probability in
particular regions of phase space where the $\nu_\mu \to \nu_e$
transition becomes resonant in matter.  Depending on the sign of
$\Delta m^2$, these effects occur either for neutrinos or for
anti-neutrinos only.  By comparing the neutrino and anti-neutrino
distributions, the sign of $\Delta m^2$, and therefore the neutrino
mass hierarchy, could be determined if a signal were observed. The
size of this effect would measure the admixture of electron neutrinos.

In this work, the sensitivity to matter effects of a massive
magnetized detector for atmospheric neutrinos is discussed in the
framework of a three flavour scenario with one mass scale dominance
for atmospheric neutrinos (i.e.  $\Delta m^2=m^2_3 - m^2_{1,2}$).  The
full simulation of the MONOLITH detector \cite{Monolith} is used for a
realistic description of resolution and reconstruction effects.  About
200~kty of MONOLITH exposure (6~y) will be sufficient to explore
regions not yet ruled out by the bound on $\nu_e \to \nu_x$
oscillations derived by CHOOZ results \cite{CHOOZ}. The potential
sensitivity of a detector (or array of detectors) with masses
significantly larger than MONOLITH 
will be described.

\section{Three neutrino oscillations with one mass scale dominance
for atmospheric neutrinos}

The current phenomenology of neutrino oscillation cannot be reconciled
with a three-neutrino oscillation scenario, unless one of the
experimental evidences for neutrino flavour conversion is
disregarded\footnote{Technically oscillations have not been observed
yet.}.  In the ``one mass scale dominance'' for atmospheric neutrino
scenario \cite{One}, the choice is made to disregard the LSND result
\cite{LSND} and describe the $\nu_\mu$ disappearance observed with 
atmospheric neutrinos as oscillations of $\nu_\mu$ into $\nu_\tau$, 
which are (almost) pure admixtures of two mass eigenstates (say
$\nu_2$ and $\nu_3$) with $\dm_{23} = \dm_{Atm}$. Solar neutrino
involve oscillations between $\nu_1$ and $\nu_2$ with $\dm_{12} =
\dm_{Sun}$ $<< \dm_{Atm}$.  

In this scheme, $\dm_{Sun}$ is too small to affect oscillation of
atmospheric neutrinos (i.e. $\Delta m^2_{Sun} L/E << 1$) and can be
approximated to zero. In this limit and up to terms proportional to
the identity, the Hamiltonian for the time evolution of neutrinos
propagating in matter of constant density reads:
\begin{equation}
\label{eq1}
H=\frac{1}{2 E}\left\{U \left(\begin{array}{c c c}
0&&\\&0&\\&&\dm\end{array}\right) U^+ + \left(\begin{array}{c c
c}2E\sqrt{2} G_F N_e&&\\&0&\\&&0\end{array}\right) \right\},
\end{equation}
where  $\dm = \dm_{13} \simeq \dm_{23}$, $\sqrt{2} G_F N_e$ is the
effective potential for the scattering of electron neutrinos off the
external field due to electrons of volume density $N_e$ and $U$ is the
flavour mixing matrix in vacuum\footnote{The corresponding
Hamiltonian for anti-neutrinos is obtained by changing $U \rightarrow
U^*$ and the sign of the effective potential.}. 

For a given baseline $L$ and constant matter density,
\begin{equation}
\nu(L) = S(L) \nu(0)
\end{equation}
with
\begin{equation}
S(L)=\tilde{U} \left(\begin{array}{c c c} 0&&\\&e^{-i\frac{\dtm_2^2}{2
E}L}&\\ &&e^{-i\frac{\dtm_3^2}{2 E}L} \end{array}\right) \tilde{U}^{+},
\end{equation}
where $\dtm^2$ describes the energy-level spacings in matter and
$\tilde{U}$ the effective mixings.  The transition amplitudes for
$\nu_\alpha \to \nu_\beta$ are given by $S$ matrix elements:
\begin{equation} 
\label{eq2}
A(\alpha\to\beta;L) = S(L)_{\beta \alpha},
\end{equation}
from which the transition probabilities may be calculated.  The exact
expressions for the effective mass differences and mixings in matter
of constant density after diagonalization of the Hamiltonian of
equation (\ref{eq1}) are given in \cite{Banul}.

Equation (\ref{eq1}) implies that the (1,2) sector is inoperative in
atmospheric neutrinos, the CP phase of the mixing matrix is
unobservable and the oscillations are entirely described by three
parameters: $\Theta_{13}$, $\Theta_{23}$ and $\dm$.  
The parameter $\Theta_{13}$ measures the admixture of
electron neutrinos to atmospheric neutrino oscillations and is related 
to $U_{e3}$ of the mixing matrix by
$|U_{e3}|^2=\sin^2\Theta_{13}$. This parameter is bound to be small by
CHOOZ results ($\sin^2 2\Theta_{13} \leq 0.1$ at 90\%
C.L. \cite{CHOOZ} for large $\Delta m^2$ ), but the (1,3) transition
can become resonant in matter and sizeably modify the oscillation
probabilities of electron and muon neutrinos. 

In a medium of constant density, the resonant energy and the resonance
width are given by
\begin{eqnarray}
  E_R & = & \pm \cos 2\Theta_{13} \frac{\Delta m^2}
{2\sqrt{2}G_F N_e}, \label{er} \\ \Gamma_R & = & 2\sin 2\Theta_{13}
\frac{\Delta m^2} {2\sqrt{2}G_F N_e}, \label{gr}
\end{eqnarray}
where the $+$ (-) sign in eq. (\ref{er}) is referred to neutrinos
(anti-neutrinos). Thus the resonance occurs either for neutrinos or
for anti-neutrinos only, 
depending on the sign of $\dm$. In the limit of small mixing angles,
the resonance width shrinks like $\sin 2\Theta_{13}$ and the effect
will eventually become unobservable when sharper than the experiment
resolution.

On the other hand, the resonant energy is practically independent of
$\Theta_{13}$ and only depends on the mass difference of neutrino
states and on the medium density. Thus, the position of the resonance
can be predicted accurately when $\dm$ is known. For $\dm = 0.003$
eV$^2$, the resonant energy is around 3~GeV, 7~GeV and 10~GeV in the
Earth core, mantle and external mantle respectively, well within the
MONOLITH acceptance (see below).  These values give only a
qualitative indication of the energy region of interest for matter
effects with atmospheric neutrinos. A more precise prediction of the
effects requires that the exact density profile along the neutrino
path be considered.

\begin{figure}
\begin{center}
\mbox{\epsfig{file=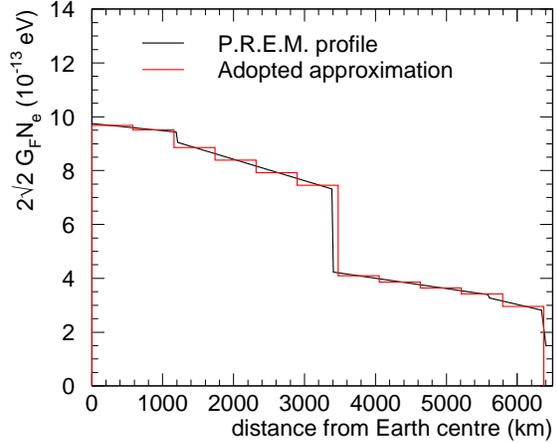,width=0.95\linewidth}}
\end{center}
\caption{Electron density profile scaled by $2\sqrt{2} G_F$ in the
Earth according to the Preliminary Reference Earth Model \cite{PREM}
and the discrete approximation adopted in this study.}
\label{PREFIG}
\end{figure}

The explicit computation of transition probabilities has been
addressed numerically.  The Earth density profile of the Preliminary
Reference Earth Model \cite{PREM} has been approximated with eleven
shells of constant density (see Fig.  \ref{PREFIG}) and neutrino
propagation has been computed from the (time ordered) product of
evolution operators in matter of constant density:
\begin{eqnarray}
\nonumber
\nu(L) & = & S(L) \nu(0) \\ & = & S_n(L_n)
S_{n-1}(L_{n-1})...~S_1(L_1) \nu(0).
\end{eqnarray}
where $S_i(L_i)$ describes the neutrino propagation along the path
$L_i$ in the $i$-th shell and $n$ is the total number of shells
crossed, which both depend on the neutrino angle.

\begin{figure}
\begin{center}
\vspace{-.3cm}
\mbox{\epsfig{file=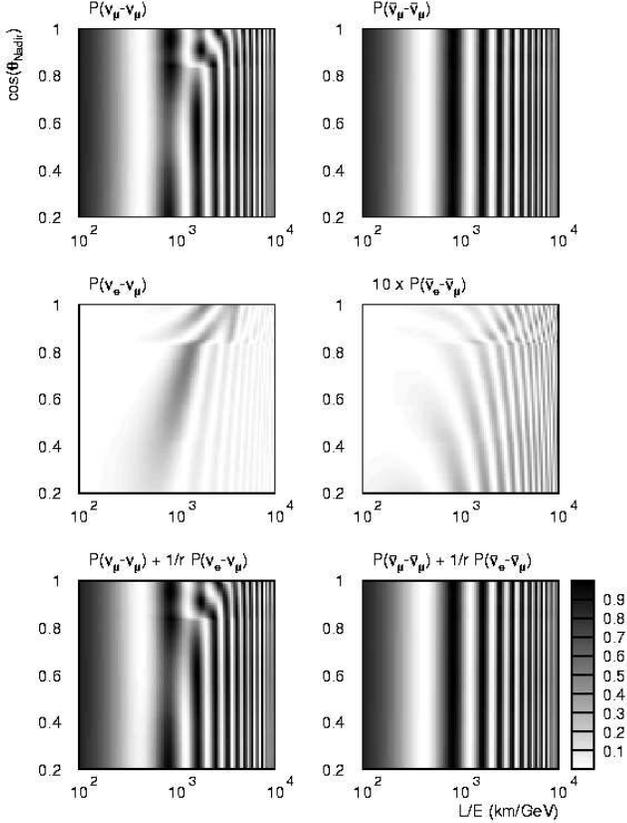,width=0.99\linewidth}}
\end{center}
\caption{In the upper and middle panels, the conversion
probabilities for atmospheric neutrinos crossing the Earth are shown
as a function of $L/E$ and $\cos\theta_{Nadir}$ for $\dm =
+0.003$~eV$^2$ and $\sin^2(2\Theta_{13})=0.1$.  The probability for
$\overline{\nu}_e \to \overline{\nu}_\mu$ conversion is amplified by
one order of magnitude to make it visible on the same scale as the
other conversions. The lower panels shows the probability of observing
muon neutrinos considering both $\nu_\mu \to \nu_\mu$ and $\nu_e \to
\nu_\mu$ conversions and the neutrino fluxes.}
\label{probosc}
\end{figure}

The $\nu_\mu \to \nu_\mu$ and $\nu_e \to \nu_\mu$ conversion
probabilities as a function of $L/E$ and $\cos \theta_{Nadir}$ are
shown in figure \ref{probosc} for $\dm = +0.003$~eV$^2$ and
$\sin^2 2\Theta_{13}=0.1$. These values of oscillation parameters 
sit at the border of the 90\% C.L. region excluded by CHOOZ. As a 
positive sign for $\dm$  was chosen, the $L/E$ pattern typical of
vacuum oscillation is practically unaffected by matter effects for
anti-neutrinos, while it is clearly modified for neutrinos. 
For neutrinos crossing the Earth's core (i.e.  $\cos \theta_{Nadir}
>0.85$), sharp structures appear in the oscillation pattern. 
For lower values of $\cos\theta_{Nadir}$, the distortion with respect
to the vacuum $L/E$ pattern for muon neutrinos is broader and most
relevant around the first maximum, with a suppression of about 50\%,
and the second minimum of the conversion probability.  
Electron neutrino oscillations into muon neutrinos are correspondingly 
enabled in the same region of the $L/E$ and $\cos\theta_{Nadir}$
plane. The energy and angular regions of interest for these effects
range from 4 GeV to 15~GeV and from $40^o$ to $60^o$, corresponding to
a baseline of about 9000~km.

For smaller values of the mixing angle $\Theta_{13}$, the energy and
angular positions of this distortion are almost independent of $\sin^2
2\Theta_{13}$, but the distortion becomes sharper, in agreement with
the expectations for neutrino propagation through a constant density
medium (like the Earth's mantle in first approximation). 

The MONOLITH detector (see below) is optimized to reconstruct muon
neutrinos and anti-neutrinos, the expected distribution of which is a
combination of the conversion probabilities for 
$\nu_\mu \to \nu_\mu$ ($P_{\mu \mu}$) and $\nu_e \to \nu_\mu$ ($P_{e
\mu}$) weighted by the corresponding neutrino fluxes. The probability 
of observing muon neutrinos is thus given by $P_{\mu \mu} + 1/r 
P_{e\mu}$, where $r$ is the $\nu_\mu$ to $\nu_e$ flux ratio. Over the
region where relevant matter effects occur, $r$ ranges between 2 and
3 and the disappearance of muon neutrinos through matter effects is only
partly compensated by $\nu_e \to \nu_\mu$ transitions. 
This is shown in the lower panels of the figure for both neutrinos and
anti-neutrinos: the flux suppression with respect to pure
$\nu_\mu$--$\nu_\tau$ oscillations is still about 30\% around the
first maximum of the reappearance probability. The detailed simulation
is addressed in the following. 

\section{Experiment simulation}


The differential distribution of the atmospheric neutrino fluxes at
Gran Sasso has been generated according to the predictions of the 1-D
Bartol model \cite{Bar96}.  Neutrino interactions have been calculated
with GRV94 parton distributions \cite{GRV94} with explicit inclusion
of the contribution of quasi-elastic scattering and of single pion
production to the neutrino cross-sections \cite{LLS95}.

\begin{figure}
\begin{center}
\vspace{.4cm}
\epsfig{file=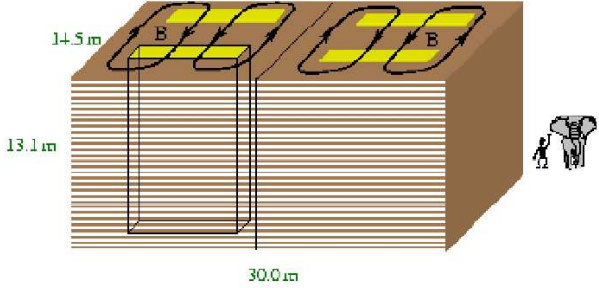, width=0.95\linewidth}
\caption{Schematic view of the MONOLITH detector.  The
arrangement of the magnetic field is also shown.}
\label{fig:module}
\end{center}
\end{figure}

Events have been processed through the full simulation and
reconstruction programs of the MONOLITH detector described in
\cite{Monolith}.  The proposed detector for MONOLITH
is a massive tracking calorimeter with a coarse structure and intense
magnetic field. The detector has a large modular structure (figure
\ref{fig:module}). One module consists in a stack of 120 horizontal 8
cm thick iron planes with a surface area of $14.5\times 15\ {\rm
m^2}$, interleaved with 2 cm planes of sensitive elements. The total
mass of the detector for two modules is about 34~kt. The magnetic
field configuration is also shown in figure \ref{fig:module}; iron
plates are magnetized at a magnetic induction of $\approx 1.3$~T.  

The active elements (Resistive plate chambers
with glass electrodes \cite{Gust??}) provide two coordinates 
with a pitch of 3 cm and a time resolution of order 1~ns

After reconstruction, some minimal requests have been applied to
select a pure sample of muon neutrino charged current interactions:
\begin{itemize}
\item
a muon candidate firing at least 7 layers is required;
\item 
the muon energy must exceed 1.5 GeV;
\item
the event is required to be either fully contained in a fiducial
volume defined by requiring no hits in the first/last 4 layers and no
hits in the first/last 10 strips in $X$ and $Y$, or to have a single
outgoing track (muon) with a reconstructed range greater than 4
metres.
\item 
the $L/E$ resolution of the event, determined from the angular and
energy resolution and estimated from the event kinematic, is required
to be better than 50\% FWHM. 
\end{itemize}
These selections guarantee a relatively high efficiency for an
adequate resolution in the reconstruction of the neutrino 
energy and direction in the region of interest for matter
effects. They result in an effective threshold around 3~GeV
of neutrino energy and in a fairly constant efficiency between 50\%
and 55\% from 5 GeV onwards. 
The neutrino angle is estimated from the muon direction with a
resolution of about 15 degrees at 3 GeV and then improving with a
$E_\nu^{-0.5}$ power law.   
The neutrino energy, estimated from the muon and hadron energy, shows 
a resolution of about 15-20\% in the energy region where resonant
effects are expected (4-15 GeV), smoothly broadening to 25\% at very 
high energies, reflecting the worsening of the muon momentum
resolution from track curvature.  Charge assignment is correct in more
than 95\% of the cases, without additional requests on the track fit
quality. The background due to internal and external sources is
estimated to be negligible. More details on the MONOLITH performance 
can be found in \cite{Monolith}. 

Only internal events are selected and, in the energy region of
interest for resonant effects, about 75\% of these are fully contained 
in the detector. External events (up-going muons from neutrino
interactions in the rock below the detector) are of little use in this
study. Their average neutrino energy is much higher than the resonant
energy and the resolution in the reconstruction of the neutrino energy
is insufficient to resolve the typical resonance width.

\section{Analysis and results}
\label{analysis} 

\begin{figure}[t]
\begin{center}
\mbox{\epsfig{file=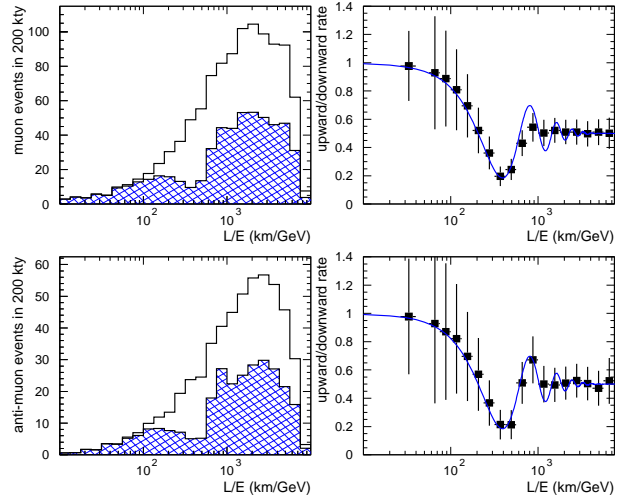,width=0.95\linewidth}}
\end{center}
\caption{
    Left: $L/E$ spectra for up-going (hatched histogram) and down-going
    muon neutrino (top) and muon  anti-neutrino (bottom) events
    for three neutrino oscillations  $\dm = +0.003$~eV$^2$, $\sin^2
    2\Theta_{23} = 1.$ and $\sin^2 2\Theta_{13} = 0.1$. 
    Right: Ratio of the up-going to down-going $L/E$ distributions. The
    curve represents the expectations for pure $\nu_\mu \to \nu_\tau$
    oscillations. 
    The experiment has been simulated with high statistics. Event
    rates and error bars are normalized to 6 y of MONOLITH exposure
    (200 kty).}  
\label{LEP}
\end{figure}

\begin{figure}
\begin{center}
\mbox{\epsfig{file=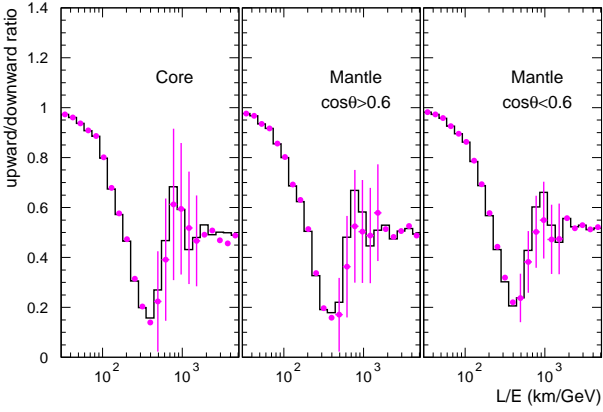,width=0.95\linewidth}}
\end{center}
\caption{
    $L/E$ spectra for up-going neutrinos crossing the Earth's core
    (left) and mantle (middle and right) assuming $\dm = +0.003$~eV$^2$, 
    $\sin^2 2\Theta_{23} = 1.$ and $\sin^2 2\Theta_{13} = 0.1$ (dots) 
    compared to the expectations for pure $\nu_\mu \to \nu_\tau$ 
    oscillations (histogram). 
    The experiment has been simulated with high statistics. The 
    error bars, shown only in the region where matter effects are
    sizeable, are normalized to 6 y of MONOLITH exposure (200 kty).}  
\label{coma}
\end{figure}

Although matter effects make the $L/E$ variable not sufficient for
oscillation studies, an optimized analysis can be performed in the
$L/E$ and $\cos \theta_{Nadir}$ plane. The energy of the resonance and
the oscillation phase in vacuum scale both with $\dm$ and the main
distortion to the vacuum oscillation pattern always occurs around the
first maximum of the reappearance probability. As visible in figure 
\ref{probosc}, the $L/E$ pattern is only marginally modified over the
first oscillation period by matter effects.  
This difference is even less evident when resolution and detector
efficiency effects are considered (see Fig. \ref{LEP}). For
$\dm=+0.003$~eV$^2$ and the largest values of $\sin^22\Theta_{13}$
allowed by CHOOZ results, the distortion on the $L/E$ pattern
(integrated over all neutrino baselines) is barely visible and located
beyond the first minimum in the survival probability. As demonstarated
in figure \ref{coma}, this distortion is mainly due to neutrinos
crossing the Earth's mantle and in any case no distortions are
predicted over the first half-period of oscillation. It follows that 
the measurement of the oscillation frequency from the position of the
first minimum in the $L/E$ pattern is marginally affected by the
presence of matter effects. The analysis of the $L/E$ pattern,
assuming that no matter effects are present, will therefore precisely
measure $\dm$ and tell where such effects have to be searched for in a
two dimensional analysis ({\it bin optimization}).

The analysis of simulated data has been based on a binned maximum
likelihood fit. The expected $L/E$ distributions for neutrino and
anti-neutrino events have been fitted to the reconstructed data from
simulation in five $\cos\theta_{Nadir}$ bins of width 0.2. The binning
in $L/E$ has been tuned according to the estimated value of $\dm$ from
a preliminary fit of the $L/E$ distribution. 

The expected number of events in each bin is a function of the
oscillation parameters ($\Theta_{13}, \Theta_{23}, \dm)$ and of the
neutrino fluxes of all the flavours. In the real experiment, 
down-going neutrinos will constitute a reference sample of
``unoscillated'' neutrinos to check the muon neutrino and
anti-neutrino rates. The rate of muon neutrinos originating from
electron neutrino conversions will have to be predicted by
flux calculations of electron neutrinos. These predictions will be
constrained both by the measurement of the correlated rate of
down-going muon neutrinos in the same detector and by the direct
measurement of the electron neutrino to muon neutrino flux ratio 
in other experiments (Super-Kamiokande). 
A detailed study to quantify the related amount of systematic
uncertainty has not been performed and an overall uncertainty of 10\%
in the predicted rates of neutrinos and anti-neutrinos has been
assumed for a preliminary estimate of the experiment sensitivity.  

The likelihood function has been defined as:
\begin{equation}
\label{like} 
\ln {\cal L} = \ln \prod_{i,q}
    \left[ 
    \frac{e^{-A_q\mu_{i,q}} (A_q\mu_{i,q})^{U_{i,q}}}{U_{i,q}!}
    \right]
    - \sum_q \frac{1}{2}\frac{(A_q-1)^2}{\sigma_A^2}
\end{equation}
where the subscripts $i$ and $q$ are referred to bins in the 
($\cos\theta_{Nadir}$, $L/E$) plane and of muon charge respectively; 
$U_{i,q}$ and $A_q \mu_{i,q}$ are the observed and expected number of
up-going neutrino events in the ${i,q}$-th bin and the last term in
the likelihood accounts for flux, cross-section and selection
efficiency uncertainties for neutrinos and anti-neutrinos
separately.

Two different analyses have been performed to determine the
sensitivity to the mixing parameter and to the sign of $\dm$. 

\subsection{Sensitivity to $\sin^2 2\Theta_{13} $} 

In this analysis it has been assumed that the mixing in the (2,3)
sector be maximal ($\sin^2 2\Theta_{23}=1$). This corresponds to the
best current estimate from Super-Kamiokande data. Individual 
experiments of high statistics have been generated for different
values of $\sin^2 2\Theta_{13} $ and $\dm$ and tested with a
likelihood ratio analysis against the pure $\nu_\mu / \nu_\tau$
oscillation scenario (i.e. $\sin^2 2 \Theta_{13} = 0$), assumed as
{\it null hypothesis}. From this procedure, the values of the
oscillation parameters that would give an experimental result
incompatible with the null hypothesis have been determined.   
The likelihood ratio is defined by: 
\begin{equation}
 \lambda = \frac{{\cal L}
 (A_q, \dm,\sin^2 2\Theta_{23} ,\sin^2 2\Theta_{13} | 
                                  \sin^2 2\Theta_{13}=0)}  
                {{\cal L} 
 (A_q, \dm,\sin^2 2\Theta_{23},\sin^2 2\Theta_{13})},
\end{equation}
representing the ratio of the maximum likelihood over the parameter
space of the null hypothesis to the maximum likelihood over the entire
parameter space. The test statistics $-2\ln \lambda$ asymptotically
behaves as a $\chi^2(1)$ distribution, whence the iso-probability
curves have been calculated. The expected and observed rates of events
in the likelihood function of equation (\ref{like}) have been
normalized to represent experiments of finite statistics. 

\begin{figure}
\begin{center}
\mbox{\epsfig{file=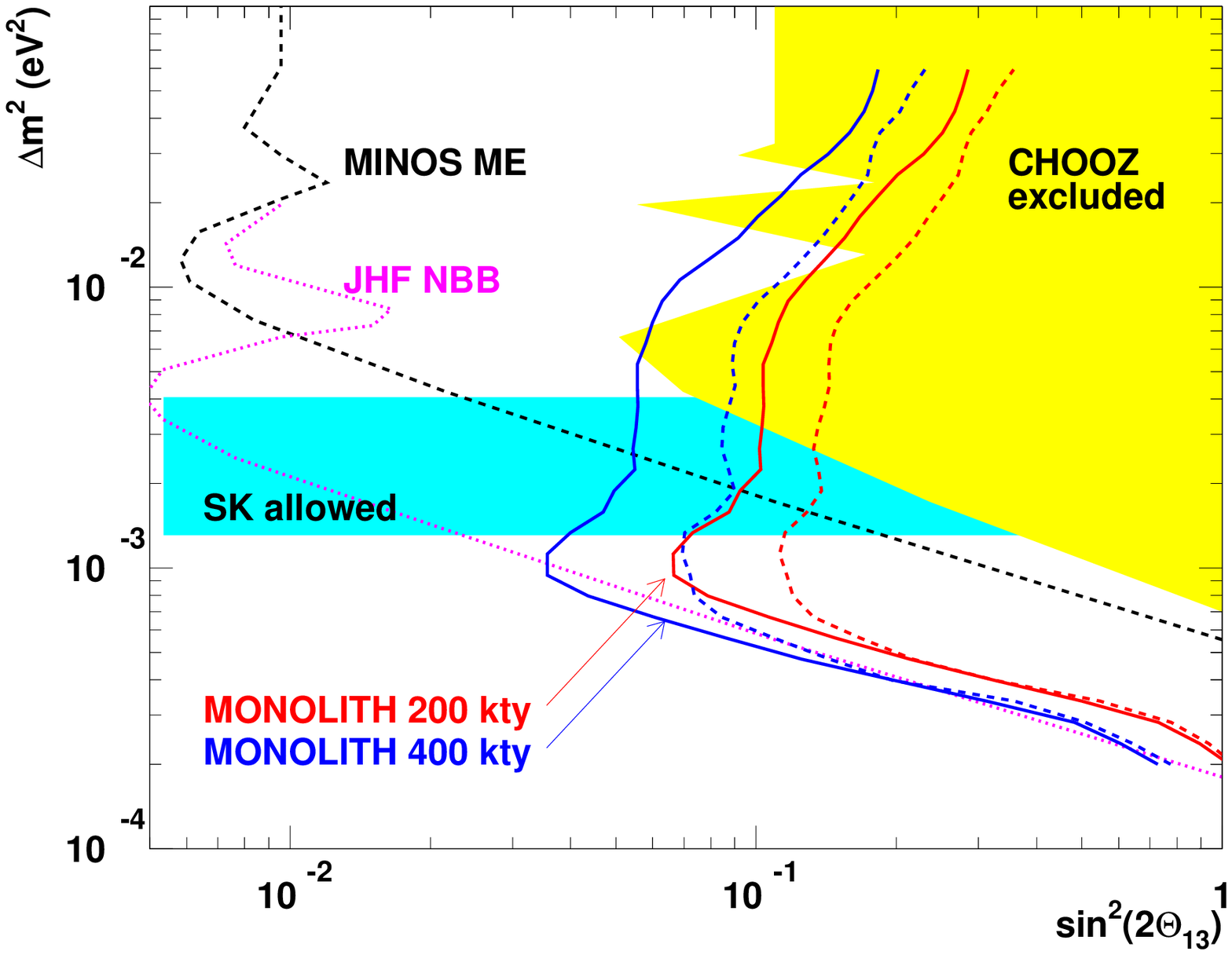, width=0.95\linewidth}}
\end{center}
\caption{
  Exclusion regions at 90\% C.L. if no matter effects are observed in
  MONOLITH after 200~kty (6 y) and 400~kty (12 y) of exposure. 
  The curves for both positive (continuous line) and
  negative (dashed line) sign of $\dm$ are given. The regions excluded by
  CHOOZ results and allowed by Super-Kamiokande data are shown,
  together with the  expected sensitivities of MINOS (Medium Energy
  option) and JHF projects (Low Energy option). The MINOS sensitivity is 
  practically independent of the beam optics configuration in the
  region of $\dm$ allowed by Super-Kamiokande data. The JHF
  sensitivity can be better or worse than the one displayed by a
  factor of two depending on the beam optics.}
\label{sensino}
\end{figure}

The expected sensitivities (exclusion regions at 90\% C.L. if no 
effect is observed) after 200~kty of exposure (about 6 y for 34 kt) 
are shown in Fig. \ref{sensino}.
Since the sign of $\dm$ is not known {\t a priori} the curves for both
positive (continuous line) and negative (dashed line) sign of $\dm$
are given. The expected exclusion region for positive sign of $\dm$ is
about two times larger than for negative sign. This follows from the
difference in neutrino and anti-neutrino cross sections, fluxes and
selection efficiencies, which make the event rate of atmospheric
neutrinos about two times larger than the rate of anti-neutrinos.

The sensitivity achievable in a long-term run (or with a larger
detector), corresponding to an exposure of 400~kty, is also shown in
the figure. 

For comparison, the region excluded by CHOOZ results \cite{CHOOZ} and
allowed by Super-Kamiokande data \cite{SK01} are displayed. Also
displayed are the expected sensitivities of neutrino oscillation programs
with conventional neutrino beams in the short/medium term. The region
covered by MONOLITH will be partly accessible to the MINOS experiment 
at NuMI \cite{Para} and fully accessible to the JHF beam to
Super-Kamiokande \cite{JHF}. The latter is expected to achieve this
sensitivity around 2012 after 5~y of operation.  

\subsection{Sensitivity to the sign of $\dm$} 

\begin{figure}
\begin{center}
\mbox{\epsfig{file=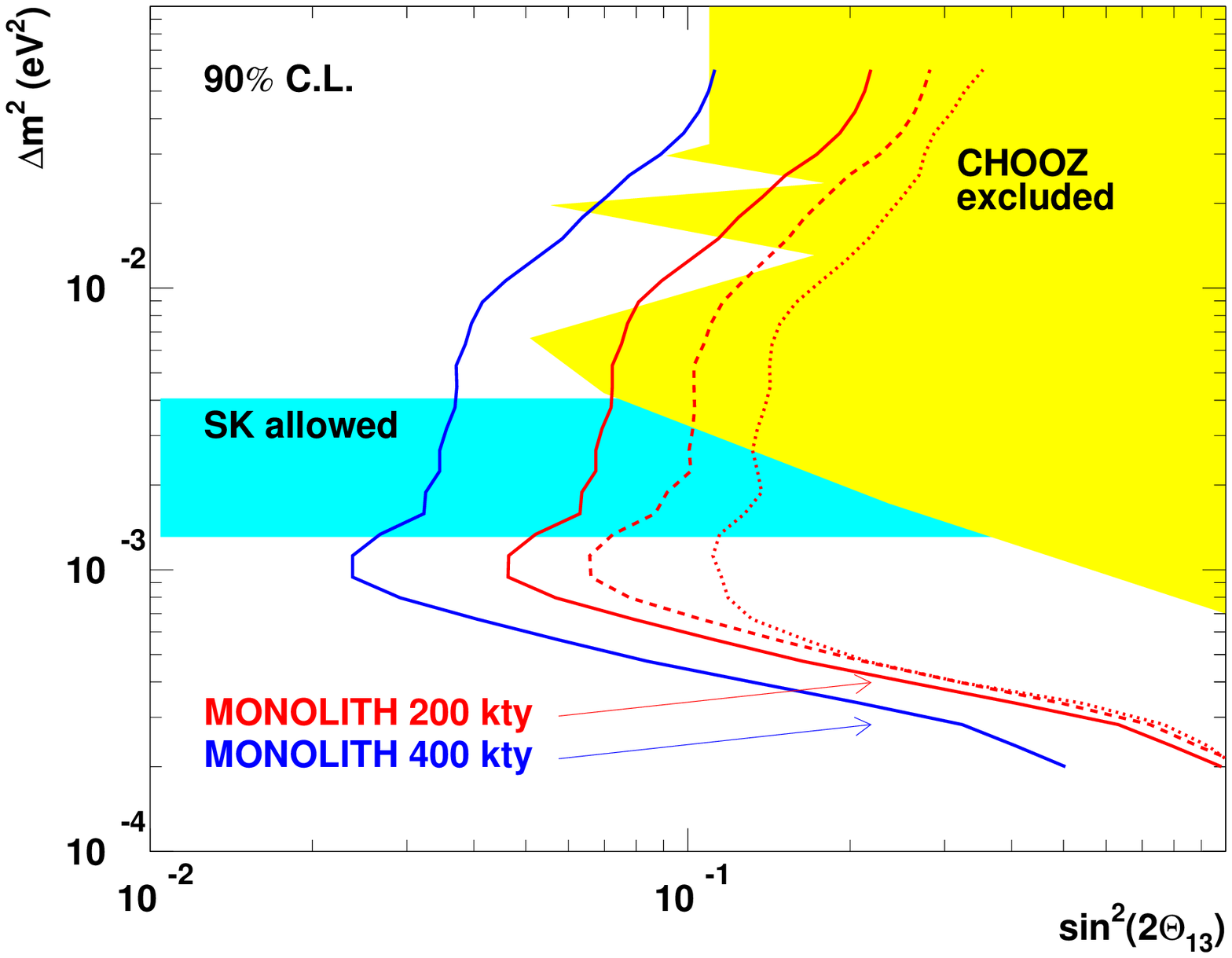,width=0.95\linewidth}}
\mbox{\epsfig{file=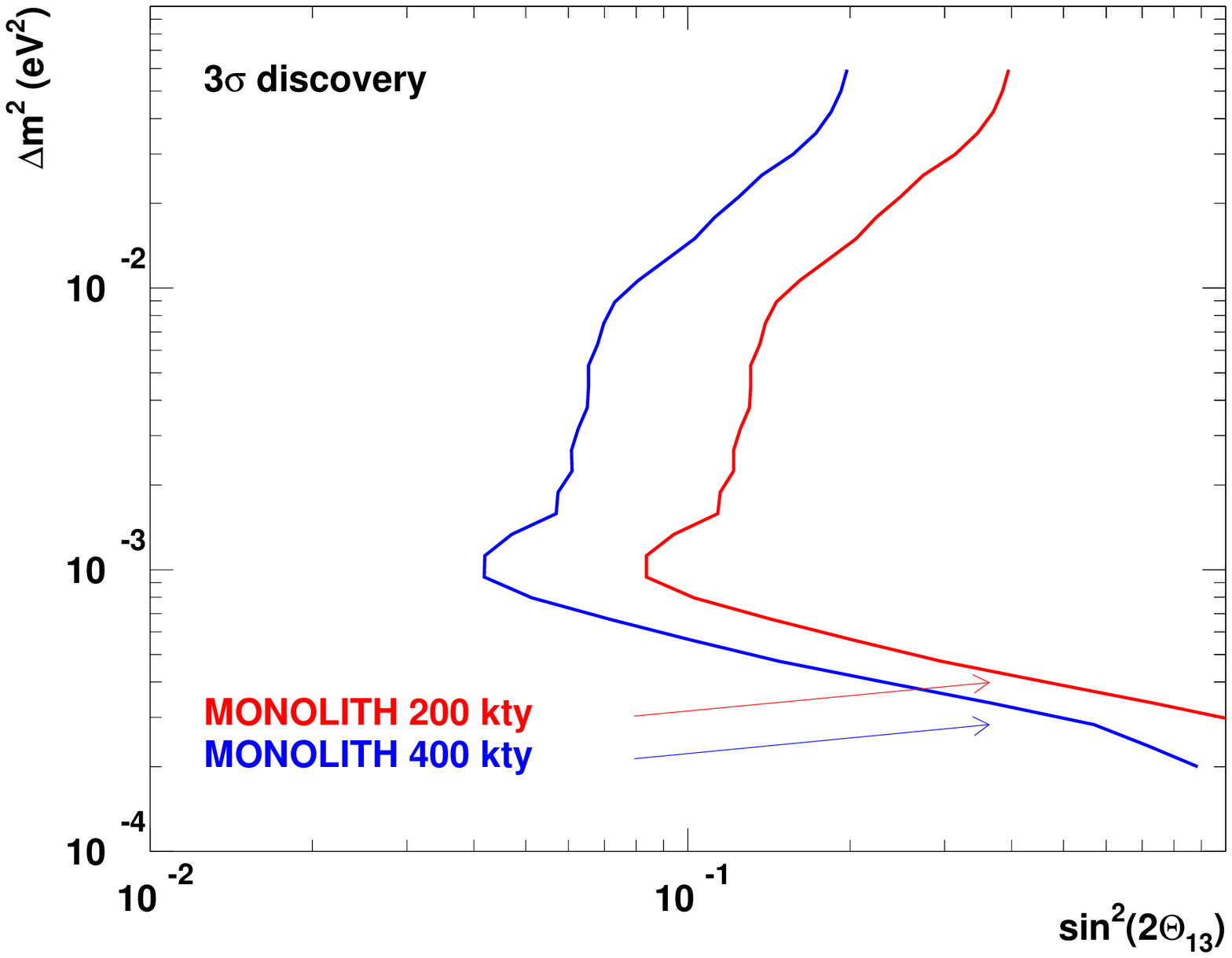,width=0.95\linewidth}}
\end{center}
\caption{
  Upper panel: The continuous lines define the boundary of the regions of
  oscillation parameters in the three generation scenario over which
  the sign of $\Delta m^2$ can be determined at the 90\% C.L.  
  after MONOLITH exposures of 200~kty (6 y) and 400~kty (12 y), assuming
  that $\sin^2 2\Theta_{13}$ be known with 30\% accuracy.
  For comparison, the broken lines show the regions over which the 
  sign of $\dm$ can be determined after 200~kty exposure 
  assuming no prior knowledge on $\sin^2 2\Theta_{13}$ regardless
  of the sign of $\dm$ (dotted) and for positive $\dm$ (dashed). 
  The regions excluded by CHOOZ results at 90\% C.L. and allowed by
  Super-Kamiokande at 90\% C.L. data are also shown. 
  Lower panel: Regions of oscillation parameters over which
  the sign of $\Delta m^2$ can be determined at the 3$\sigma$ level
  after MONOLITH exposures of 200~kty (6 y) and 400~kty (12 y), assuming
  that $\sin^2 2\Theta_{13}$ be known with 30\% accuracy.} 
\label{matm}
\end{figure}

Experiments at NuMI and JHF search directly for the sub-dominant
$\nu_\mu \to \nu_e$ transition and have too short baselines to exploit
matter effects. Thus, they cannot measure the sign of $\Delta m^2$.   

With atmospheric neutrinos, if an effect is observed, the sign of
$\dm$ can also be determined. Assuming no prior knowledge on $\sin^2
2\Theta_{13}$,  this happens at the 90\% C.L. on the right-hand side
of the dashed curves of figure \ref{sensino}, irrespectively of the
sign of $\dm$, and on the right-hand side of the continuous curves, if
$\dm$ is positive.   

However, as discussed in the previous analysis, provided that the
mixing  $\sin^2 2\Theta_{13}$ be large enough, there is a fair chance
that it will be measured in the next generation of long base-lines
experiments. In that case, this information can be used to constrain
the maximum likelihood fit and and determine the sign of $\dm$ from
the muon/anti-muon rate asymmetry. 

The likelihood function of eq. (\ref{like}) can be modified as:
\begin{equation}
\ln {\cal L}' = \ln {\cal L} 
-\frac{1}{2}\frac{(\sin^2 2\Theta_{13}-<\sin^2 2\Theta_{13}>)^2}{\sigma_{13}^2}
\end{equation}
In this analysis,  $\sin^2 2\Theta_{23}$ has been assumed to be
maximal and  $\sin^2 2\Theta_{13}$ has been assumed to be known with
an accuracy of 30\%, representing the observation of a 3$\sigma$
effect at accelerator beams. Similarly to the analysis outlined in the
previous section, individual experiments have been generated  for
different values of $\sin^2 2\Theta_{13} $ and $\dm$. For each
(absolute) value of $\dm$, the minimum value of $\sin^2 2\Theta_{13}$
for which the best fit to data with the wrong sign of $\dm$ is
incompatible with data has been determined.  

Figure \ref{matm} (upper panel) shows the expected region of
oscillation parameters over which the sign of $\Delta m^2$ can be
determined at the 90\% C.L. after 200~kty of MONOLITH exposure (i.e. 
the wrong sign of $\Delta m^2$ is excluded at the 90\% C.L.). The
sensitivity achievable in a long-term run, corresponding to an
exposure of 400~kty, is also shown. For comparison, the region
excluded by CHOOZ \cite{CHOOZ} and the region allowed by
Super-Kamiokande at 90\% C.L. are displayed.  On the same figure, the
regions over which the the sign of $\dm$ can be determined after 200
kty of exposure without any prior knowledge of the (1,3) mixing are
shown for comparison. The  3$\sigma$ discovery limit for the sign of 
$\Delta m^2$ is shown in the lower panel of the figure.

As expected, the additional constraint on the mixing parameter
widen the sensitivity region and an exposure of 200 kty would be
sufficient to evade the CHOOZ exclusion region. This is because
in this case one has only to discriminate a positive from a negative
asymmetry, while assuming no prior knowledge of the (1,3) mixing, 
an asymmetry of any sign should be discriminated from the no-asymmetry 
case.

\section{Systematic uncertainties}

Several sources of systematic uncertainties might affect the precision
of this measurement.

The likelihood ratio method described by eq. (9) implies that no prior 
knowledge on the values of $\dm$ and $\sin^2 2\Theta_{23}$ is assumed. 
These parameters can be measured in MONOLITH with an accuracy of about
5\% with 200~kty exposure and reaching 1\% for very large exposures. 
Since this would represent a substantial improvement with respect to
the present knowledge, these parameters are not constrained in the
fitting procedure.

The limited knowledge of the electron density profile in the Earth is
also a source of systematic uncertainty. However, the sensitivity to 
matter effects with atmospheric neutrinos is dominated by the resonant
behaviour of neutrinos crossing the bulk of the Earth's mantle, whose 
density and chemical composition are known with sufficient precision.
The matter density of the mantle is strongly constrained by the
knowledge of the Earth's mass and moment of inertia, combined with
minimal information from seismic wave measurements of the density
profile. The electron density is related to the matter density by the
$Z/A$ ratio, which is very nearly 1/2 for all reasonable chemical
compositions of the Earth's mantle.

As discussed in section \ref{analysis}, the effects related to the
limited knowledge of neutrinos and anti-neutrino fluxes and cross
sections have been preliminary accounted for in this analysis by
assigning an uncertainty of 10\% to the predicted rates for neutrinos
and anti-neutrinos separately. For an exact knowledge of these rates,
the sensitivity regions shown in figure \ref{sensino} would have to be 
extended by about 10\%. 


In a real experiment, the sample of down-going neutrino events can be 
exploited to constrain the neutrino and anti-neutrino event rate
predictions. The precision of this procedure might be expected to
scale with the square root of the exposure and will not be a
fundamental limitation to the sensitivity to matter effects with
atmospheric neutrinos. This is further discussed in the next section. 

\section{Very large exposures}

The dependence of the experiment sensitivity on the exposure has  been
studied. The number of years of MONOLITH exposure has been  increased
up to represent a collected statistics of 2~Mty. Since selection and
reconstruction efficiencies are the ones of MONOLITH, this simulation
describes a detector (or array of detectors) of modular structure,
with each single module comparable to MONOLITH in  acceptance and
performance.

\begin{figure}
\begin{center}
\mbox{\epsfig{file=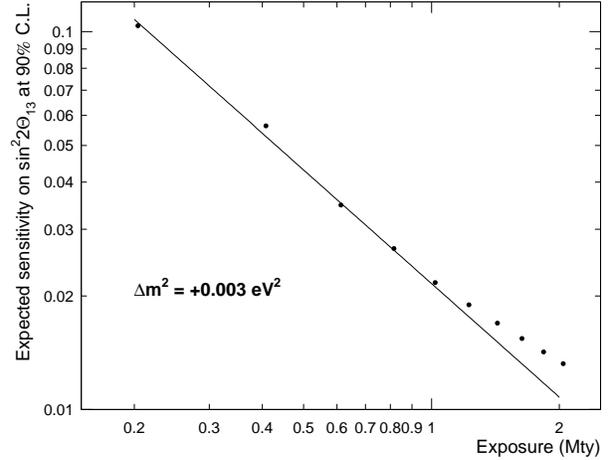, width=0.95\linewidth}}
\end{center}
\caption{ Expected exclusion regions at 90\% C.L. as a function of the
exposure, if  no matter  effects are observed  in MONOLITH.  Data have
been simulated with $\dm =  +0.003$ eV$^2$.  The continuous line shows
the  expected scaling behaviour,  when the  sensitivity is  limited by
statistical fluctuations (see text).}
\label{long}
\end{figure}

Figure  \ref{long}  shows  the   expected  sensitivity  to  $\sin^2  2
\Theta_{13}$ as a  function of the exposure for $\dm  = + 0.003$.  For
negative sign of  $\dm$ the sensitivity is about  two times worse.  

In this study, the systematic uncertainty related to the knowledge of
the expected neutrino rates has been assumed to be 10\% up to 400~kty 
and scaled with the square root of the exposure at the other points.
Indeed, in a real experiment, down-going neutrino events will provide
a reference sample of ``unoscillated'' neutrinos to check the expected
muon neutrinos and anti-neutrinos rates with statistically increasing
precision. 

The sensitivity to matter effects would be expected to improve
linearly with the exposure, if it were limited only by statistical
fluctuations. In the region around the resonance, matter
effects determine a $\nu/\overline{\nu}$ asymmetric variation in
the event rate with respect to pure $\nu_\mu/\nu_\tau$
oscillations, that scales as the width of the resonance, i.e.
$\sin 2\Theta_{13} = \sqrt{\sin^2 2\Theta_{13}}$. This has to
exceed the statistical fluctuations in the same region, which
decrease as $1/\sqrt{T}$, where $T$ is the total exposure. Thus the
expected exclusion limit on $\sin^2 2 \Theta_{13}$ should scale as
$1/T$.

As shown  in the figure, this scaling  law holds up to  an exposure of
around 1~Mty, corresponding to a sensitivity around 0.02 on the mixing
parameter. Beyond that limit, the width of the resonance in the Earth
mantle becomes comparable to the energy resolution of the experiment
and the resonance cannot be fully resolved. An improvement  in the
energy  resolution  would  thus be  necessary to further push the
sensitivity to matter effects for $\sin^2 2\Theta_{13}$ below 0.01.  

Following  the discussion  given in  \cite{Banul}, for  values  of the
mixing  parameter this low,  the observation  of the  resonance should
also  be inhibited  by other  reasons.  A sizeable  distortion of  the
oscillation pattern can  be observed only when an  overlap between the
resonance in  the effective mixing and the  oscillation factor occurs.
According  to this  argument,  in  a medium  of  constant density  the
observability  of the  resonance  requires a  minimum baseline,  whose
length scales as $1/\tan 2\Theta_{13}$.  For mixings as small as 0.01,
this baseline exceeds the Earth's diameter.

\section{Conclusions} 

A  study of  the  sensitivity  of a  massive  magnetized detector  for
atmospheric neutrinos (MONOLITH) to matter effects in the framework of
a three flavour scenario with one mass scale dominance for atmospheric
neutrinos has been reported.

A sub-dominant component of $\nu_e$ mixing could give rise to resonant
$\nu_\mu  \leftrightarrow \nu_e$  transitions in  matter,  which occur
either for neutrinos or for anti-neutrinos only, depending on the sign
of $\Delta  m^2$. This  can be exploited  in a magnetized  detector to
determine the  sign of  $\Delta m^2$ and  therefore the  neutrino mass
hierarchy. The  size of this effect  would also measure  the mixing of
electron neutrinos.

About 200~kty of MONOLITH exposure will be sufficient to explore
regions not yet ruled out by the bound on  $\nu_e \to \nu_\mu$
oscillations derived by CHOOZ results \cite{CHOOZ}. These regions 
will be partly accessible to MINOS and JHF \cite{Para,JHF}, which,
however, can not measure the sign of $\Delta m^2$.  

The sensitivity achievable in a high statistics run has also been 
discussed. The ultimate sensitivity which can be achieved with
exposures of 1-2 Mty would be $\sin^2 2 \Theta_{13} \sim
0.01$. In order to go below this limit, the energy resolution would
have to be improved. For values of $\sin^2 2 \Theta_{13}$ this low,
the observability of the resonance is also compromised by the fact
that the maximum of the effective mixing is somehow compensated by a
minimum in the oscillation factor over the Earth's dimension.

\section*{Acknowledgements}
Stefano Ragazzi is gratefully acknowledged for his encouragement, for
critical comments and suggestions during the entire development of
this study. Comments and suggestions of G.~Barenboim, J.~Bernabeu,
A.~Curioni, A.Geiser and A.~Marchionni are warmly thanked.  
Stimulating questions of F.~Dydak and J.J.~Gomez--Cadenaz are 
acknowledged. %
I am indebted to G.~Battistoni and P.~Lipari for the simulation of 
atmospheric neutrino fluxes and neutrino interactions and to several 
colleagues, who have made important contributions to the development
of the MONOLITH software, and in particular F.~Pietropaolo,
P.~Antonioli, A.~Curioni, M.~Selvi, F.~Terranova and A.~Tonazzo.

\end{document}